# New class of compounds – variators – are reprogramming substrate specificity of H4K12Ac, H4K16Ac and H4K20Ac epigenetic marks reading bromodomain of BPTF protein.

Oleksandr Ya Yakovenko[*,¶], Sreeja Leelakumari[¶], Ganna Vashchenko[└], Albert Badiong[¶] and Steven J.M. Jones[¶]

[¶]*Canada's Michael Smith Genome Sciences Centre, BC Cancer Agency, Suite 100 570 West 7th Avenue, Vancouver, British Columbia V5Z 4S6, Canada*

[└]*Department of Biochemistry and Molecular Biology, University of British Columbia, 2350 Health Sciences Mall, Canada*

[*] *Corresponding author, email ayakovenko@bcgsc.ca*

*Abstract.*

Previously reported [http://arxiv.org/abs/1506.06433] reprogramming of substrate specificity of H3K4Me3 epigenetic marks reading PHD domain of BPTF protein illustrates therapeutic potential of a new class of non-inhibitor small organic compounds – variators. Here we address the question about reproducibility of rational design of variators by reprogramming of the second epigenetic marks reading domain of BPTF protein – bromodomain. Bromodomain of BPTF binds to epigenetic marks in form of acetylated lysine of histone H4 (H4K12Ac, H4K16Ac and H4K20Ac), which physicochemical properties and binding mode differs considerably from those of methylated H3K4 marks. Thus, detailed description of computational approach for reprogramming of bromodomain substrate specificity illustrates both general and target specific attributes of computer aided variators design.

*Introduction.*

Epigenetic regulatory system is of the same importance for organism development and living as sequences of genetic programs themselves. In contrast to static genes, epigenetic regulation of their transcription is dynamic, widely adjustable and reversibly implemented via coordinated activity of histone- and DNA-modifying enzymes together with ATP-dependent chromatin remodeling complexes. The later, ATP-dependent chromatin remodeling complexes, are represented by four families (SWI/SNF, ISWI, CHD, and INO80) classified mainly depending on incorporation of SWI2/SNF2-related catalytic ATPase subunit [1]. The basic component of ISWI remodeling complex is the nucleosome remodeling factor (NURF) [2]. NURF, in turn, is composed of BPTF (bromodomain PHD finger transcription factor), SNF2L, the ATPase subunit and pRBAP46/48 [3]. Among them BPTF is the protein that is responsible for recognition of activating histone modifications to apply the remodeling activity. BPTF, in context of NUFR complex, has two domains in proximity on its C-terminus: PHD finger domain that recognizes H3K4Me3 marks (the affinity is $K_d≈230$ μM for K4me0, $K_d≈46$ μM for K4me1, $K_d≈6$ μM for K4me2 and $K_d≈3$ μM for K4me3) [4, 5] and consecutive Bromodomain for readout of acetylated histone marks (the affinity is $K_d≈69$ μM for H4K12Ac, $K_d≈99$ μM for H4K16Ac and $K_d≈130$ μM for H4K20Ac [6]). Effect of BPTF mediated chromatin remodeling is better studied in the case of H4K4me3 marks that are strongly correlated with intensive transcription of underneath genes [4, 7, 8]. However reported functional properties of the interdomain linker and biased BPTF colocalization with doubly-modified nucleosomes are indicating cooperation between the two domains and synergistic effect of simultaneously present epigenetic marks H3K4Me3 and H4K16Ac [6].

We have previously reprogrammed substrate specificity of PHD finger domain of BPTF [http://arxiv.org/abs/1506.06433]. The compounds showed promising biological effects upon artificial stabilization of PHD domain on its unnatural low methylated H3K4 marks with aid of the reprogramming compound. Here we are reprogramming substrate specificity of the second synergistic domain of BPTF – bromodomain. The bromodomain reprogramming was made so to straighten binding to unacetylated lysines of histone H4 tail which binding is not detectable normally. This is expected to indirectly enforce H3K4Me3 binding due to reported cooperative allosteric effect. The enforcement would result in better transcription of otherwise repressed proteins and this way counteract tumors strategy of dedifferentiation via epigenetic suppression caused in particular by frequent loss of functions mutations of H3K4me2/3 marks writing protein MLL2 [9, 10]. As there is little known about importance of BPTF binding to H4K16Ac marks on their own, we relayed on synergy with binding to the methylated mark and selected MLL2 deficient diffuse large B-cell lymphoma cell line SU-DHL-9 [11] the reporter cell line and MLL2 wt cell line DoHH2 [12]; as the control.

The second successive reprogramming report would illustrate systematical reproducibility of our protocol for rational design of reprogramming compounds – variators. Our protocol is built on numerous molecular dynamic simulations (MD) of triple complexes of a target protein with its unnatural substrate and small organic compound that is acting as selective molecular glue. The spatial structure of BPTF bromodomain and its binding mode with acetylated histone tail is reported [6]. It shows rather typical bromodomain binding with network of hydrogen bonds for interactions with amide group of acetylated lysine in the deep hydrophobic cavity. The most critical interactions are thus substrate readouts by sidechain of residue N148 (here and later on the numerations from report about the crystal structure is used). Another important detail about the acetylated lysine binding cavity is its lateral accessibility for water and relatively big empty overall volume. Such easy accessibility of the critical readout interactions simplify bromodomain reprogramming protocol considerably as triple complexes with reprogramming candidates can be speculated by straight application of trivial molecular docking methods.

*Methods.*

*Rational design of variators.* A protocol, similar to one described in our previous report [http://arxiv.org/abs/1506.06433], was used to identify promising candidates for the reprogramming of affinity of bromodomain of BPTF. At the first stage, the set of a priory required features of the reprogramming compound were speculated and then suitable candidates were assessed with MD simulations.

Generally speaking there are three *a priori* important features of any reprogramming compound. Firstly it should contain an 'anchoring fragment' that is responsible for binding of the compound to the target. BPTF bromodomain binds acetylated lysine histone tails with deep and narrow hydrophobic cavity formed by residues V97, A102, Y105, Y147 that opens into bulk cavern formed by W91, F93 I140 and F154 (numeration is from article about resolved 3D structure of bromodomain bound to H4K12Ac tail fragment [6]). To selectively accommodate locally hydrophilic acetylated H4K12 residue, the inner wall of the cavity is incrusted with asparagine amine N148 which forms a network of hydrogen bonds with the acetylated lysine sidechain in hydrophobic media of the binding cavity. Due to significantly smaller volume of unacetylated lysine sidechain, the hydrophobic binding cavity has enough room to accommodate variety of reasonably sized 'anchoring' fragments. Better yet, unlike PHD domain of BPTF, binding cavity of bromodomain is facing water environment from its lateral side. This makes reprogramming much easier comparing to our previous work where we have to disassemble fragment aromatic cage of H3K4Me3 binding site to get access to its inner physicochemical microenvironment. Thus we had skipped stage of virtual probe identification from our previous protocol and used molecular docking algorithm directly i.e. without any special filtration for 'allowed' anchoring fragments.

Secondly, in order to manifest reprogramming effect, the compound has to interact with the unmodified lysine to stabilize it in the active site of bromodomain of BPTF. It is thus assumed to be a 'reprogramming motif' possessing a negatively charged hydrogen bond accepting (or proton donating pattern) to strengthen the binding to positively charged $R-NH_3^+$ group in the hydrophobic environment. Besides, the compensating hydrogen bonds network should be mounted in the cavity to take care about N148 sidechain that normally serves as the selective readout of acetylated marks. Despite of relatively big volume of the binding pocket of bromodomain, accordingly to molecular dynamic simulations with trial virtual probes, there is not enough room to accommodate acidic reprogramming motifs $R-COO^-$, $R-PO_3^{2-}$ or $R-SO_4^-$ at the proper distances and orientations toward H4K12 residue. Thus, less straightforward alternatives like aliphatic amines, imines, nitriles, nitroxyles, , nitrosyles, (thio)ketones, (thio)amides etc were considered as reprogramming motifs. Although neither of considered groups carry strong negative charge to interact with charged protonated lysine, they has high dipole moments and can form strong hydrogen bonds with proton-rich sidechains. Due to diversity of the mentioned options, we didn't apply any filters prior docking run but the candidate were rejected unless they were bearing at least one possible reprogramming motif at the proper spatial position of the model complex.

The third desirable feature is that a reprogramming compound should not compete with the binding of unacetylated lysine tail to bromodomain of BPTF but stabilize the unnatural complex. The feature was estimated using molecular dynamic (MD) simulations of a triple complex of bromodomain, fragment of H4K12 unmodified histone tail and the candidate molecule. The triple complexes of a candidate bound to bromoomain of BPTF with H4K12 peptide were derived from crystal structural model of the binary complex resolved by *Ruthenburg et al.* [6] (www.pdb.org, accession code 3QZV). The reported structure was edited to replace acetylated K12 with unacetylated one, PHD domain was cut at residue L69 to reduce size of the system (*Ruthenburg et al* reports about absence of measurable cooperativity in binding to peptides) and histone tail fragment was extended by 3 residues on its C-terminus and then flanked with acetyl and aminomethyl groups on its N and C-terminuses correspondingly. The candidate compounds were simply docked with molecular docking method implemented in ICM software package [13, 14]. The entire NCI platted collection of small organic compounds (http://dtp.nci.nih.gov) was docked to the binary complex of the histone tail with BPTF and 5000 of the top-scored candidates were selected for visual inspection. The candidate compounds were manually inspected to confirm that it a) has adequate hydrophobic anchoring fragment, b) bears at least one reprogramming motif in proximity of $NH_3^+$ group of H4K12 sidechain and c) possess components to at least theoretically mount local hydrogen bonding network with important N148 residue of bromodomain of BPTF. Only those candidates that successfully passed the a), b) and c) filters were considered as promising and all those failed at least one of the filters were immediately rejected. The iterative enrichment procedure, which was described in our previous work, was used here either to consider even more promising compounds. For each promising compound; the triple complex of new candidate was created (either taken from computationally predicted binding mode or constructed manually considering insights from previously considered cases) and evaluated with MD as it is described in details in separate section below. Meanwhile MD simulations of candidate were computed, the similar structures to the analysed one were searched through entire NCI database using substructures search implemented at https://pubchem.ncbi.nlm.nih.gov/. All interesting homologues were evaluated manually, regardless of the molecular docking scores they possessed, to consider all highly similar interesting molecules that were not scored well by the molecular docking software. The stability of the triple complex was instead speculated from the behaviour of the previously simulated complexes (the amount had been increasing through the selection process). In all promising cases and the cases where it was difficult to make the conclusion due to diversity of the new candidate from all those that had been already simulated, the stability of its triple complex was evaluated with MD protocol. In every case the MD trajectories were reviewed and conclusions about the importance of particular substitutions were made to improve the consequent continuation of the visual inspection of docking results.

Finally 24 compounds that provided a stable triple-complex behavior in MD simulations were chosen for testing *in vitro* with lymphoma-derived cell lines.

*Detailed MD protocol*. Prior to the simulations every triple complex model was placed in triclinic box of simple point charge (SPC) water [15] into which 100 mM NaCl equivalent was added including neutralizing counter-ions. Periodic boundaries were applied in all directions. The N- and C-termini of bromodomain domain of BPTF protein were ionized but acetyl and aminomethyl flanked tails of H4 histone fragment were unmodified. All other amino acids were assigned their canonical state at physiological pH including ionization of H4K12 sidechain (unless the nature of reprogramming motif assumed donation of its proton to R-NH2 group of H4K12). Energy terms from the GROMOS96 43a1 parameters set[16] were applied to all molecular species in the system. All triple complexes were either taken from computationally predicted docked binding modes or constructed manually by rotation, translation and dihedral angle alternations of compounds which were optimized in vacuum at PRODRG server [17]. The bonded parameters sets of compounds were made with PRODRUG server but nonbonding parameters of compounds were intensively edited manually, especially the charge distribution which was rewritten from scratch for each compound due to known lack of accuracy in parameters sets generated by PRODRG [18].

Unlike in the case of PHD domain reprogramming where we used somewhat 'stickier' non-bonded interactions, this time a typical GROMOS96 cutoffs were used: Leonard-Jones interactions were set to 1.4 nm and Coulomb short range were set to 0.9 nm with neighbor-searching repeated for every 10 steps (each of dt=0.002 picoseconds (ps)) of the MD integrator. Long-range electrostatic interactions were modeled with the particle mesh Ewald (PME) algorithm [19]. The solvated complexes were relaxed by l-BFGS minimization [20]. Then simulating annealing [21, 22] was used to warm up the system from initial velocities assigned accordingly to Boltzmann distribution at T=10K till T=310K after 100 ps under constant volume (NVT) ensemble. A. Following the NVT warm up, 100ps of constant pressure (NPT) equilibration was performed. Complex and solvent with ions were coupled to a separate temperature coupling bathes and the temperature was maintained at T=310K. For equilibration a weak coupling method[23] was used to maintain pressure isotropically at 1.0 bar and temperature constant at 310K. All subsequent productive runs (of 12 ns each) were performed with the more accurate Nose-Hoover thermostat [24, 25] with temperature coupling time constant of 0.1ps and the Parrinello-Rahman barostat [26, 27] with pressure coupling time constant of 1.0ps under NPT ensemble. This combination of thermostat and barostat ensures that a true NPT ensemble is sampled. For visual inspections of MD trajectories VMD viewer [28] was used.

*Cell Culture and Proliferation Assays.* The DLBCL cell lines were maintained in RPMI medium 1640 (Life Technologies) supplemented with 10% (v/v) fetal bovine serum (Life Technologies) and 1% penicillin/streptomycin (Life Technologies). All cell lines were grown in a 37°C incubator with 5% $CO_2$, and humidified atmosphere. Powdered drug compounds were solubilized and maintained in the drug-carrier, dimethyl sulfoxide (DMSO), at a concentration of 10mM. In preparation for cell treatment, these drug stocks were diluted in RPMI 1640 medium. Cells to be treated were harvested and made up to a suspension with a density of $4 \times 10^5$ cells/mL. 90μL of this suspension were placed into each well of a MICROTEST™ 96-well Assay Plate, Optilux™ (BD). 10μL of the drug solution was added to each well as the treatment. Also included in each plate were a background noise control (medium only), untreated cells control, and a drug-carrier (DMSO) control. Cells are kept in a 37°C incubator with 5% $CO_2$, and a humidified atmosphere with their treatment conditions for 48 hours. Proliferation of drug-treated cells was measured by incubating them with 10% alamarBlue® (Life Technologies) for 2 hours. Fluorescent signals were later measured, the background noise subtracted, and then normalized to the drug-carrier control's signal.

*Protein Preparation.* The full-length human BPTF construct was obtained from C. David Allis of the Laboratory of Chromatin Biology, Rockefeller University [5]. Primers were designed to capture the BPTF bromodomain and PHD domain. The PHD domain is responsible for recognizing H3K4me3 marks while the bromodomain is responsible for recognizing histone H4K12Ac, H4K16Ac and H4K20Ac. The dual PHD finger-bromodomain (residues 2583-2751) from the human BPTF (gi:31322942) was cloned into a pDEST15 vector using Gateway® cloning technology allowing for an N-terminal GST tag (Life Technologies). Over-expression of the GST-BPTF dual PHD finger-bromodomain was induced in BL21-AI™ chemically competent *E. coli* cells (Life Technologies) using LB medium supplemented with 1 mM IPTG (Santa Cruz Biotechnologies) and 0.2% L-arabinose (Sigma) for 2 hours in a 37°C shaking incubator. GST-BPTF dual PHD finger-bromodomain was purified using Glutathione Sepharose 4B media (GE Healthcare Life Sciences) and dialyzed against PBS (Life Technologies) overnight.

*PullDown Assay.* The peptide pulldown experiment was performed essentially as previously described [6]. Briefly, biotinylated H4K12Ac_H4K16 (1-25 histone H4 peptide with acetyl mark at K12 only) and H4K12_H4K16Ac (1-25 histone H4 peptide with acetyl mark at K16 only) peptides were purchased from AnaSpec. These peptides were immobilized on streptavidin dynabeads (Dynabeads® M-280 Streptavidin by Invitrogen™) and were used to bait the GST-tagged BPTF into binding. Fifty microlitres of M-280 streptavidin-coupled Dynabeads® (Life Technologies) were dispensed into 1.5 mL microtubes and washed 3 x 50 μL in PBS. The beads were then incubated with C-terminally biotinylated H4K12Ac_H4K16 or H4K12_H4K16Ac peptides, 25 amino acids long, rotating for 1 hour at 4°C under saturating conditions. The beads were washed 3 x 100 μL in HBS-TD (10 mM Na-HEPES, 150 mM NaCl, 0.005% Tween-20, 2 mM DTT) and incubated with 8.2 μM of GST-BPTF along with either 82 μM of drug in a 1:100 DMSO to PBS solution or carrier alone rotating for 3 hours at 4°C. The beads were then washed 10 x 200 µL in HBS-TD. The protein was eluted using 2x LDS sample buffer (Life Technologies) and 1x reducing agent (Life Technologies) at 85ºC for 10 minutes. The eluted samples were analyzed by western blot using a GST antibody (Santa Crus Biotechnologies) in order to visualize the GST-BPTF protein. Quantification of bands was made using the Image J software.

*Results.*

*Cell-based activity assay identified highly potent chemical structures.*

As a reporter for *in vitro* screening of BPTF reprogramming compounds, *MLL2* homozygous insertion/deletion mutants SU-DHL-9 [11] cell lines were used and the control was *MLL2* wild-type cell line, DoHH2 [12]; both are diffuse large B-cell lymphomas (DLBCL). Due to limited knowledge about importance of BPTF interaction with H4K12Ac/H4K16Ac/H4K20Ac marks we relayed on reported in literature synergy between PHD and bromodomain recognition of marked nucleosomes and previously observed sensitivity of SU-DHL-9 cell line to functional state of BPTF (and lower sensitivity of the control DoHH2 cell line). The screening of the 24 computationally selected compounds revealed one especially promising scaffold with differential cytotoxicity to the reporter versus control cell lines. Due to abundance of already existing derivatives, one round of additional testing of structural homologues of the identified scaffold (with only limited MD modeling) was carried out. Approximately 20 more homologues compounds from the NCI plated collection were tested with the cell lines *in vitro*. From this second screening we identified "leading" compound (NSC689846). The "lead" had successfully passed PAINS filters for probable false-positive hits [29] as it is implemented at http://cbligand.org/PAINS/search_struct.php and its mechanism of action was confirmed rigorously with pulldown assay.

*Assessment of compounds in their ability to reprogram the affinity of bromodomain of BPTF to unacetylated H4K12 and H4K16.*

The formation of the triple complex and the ability of the compounds to alter the affinity of BPTF bromodomain toward unmodified H4K12, H4K16 and H4K20 tail peptides were confirmed with pulldown experiment.

The pulldown assay confirms the reprogramming effect of the compounds altering the stability of BPTF complexes with histone H4 tail peptides in either acetylated or unacetylated states (Fig 1). The two tested peptides have identical N-terminal histone H4 sequence (1-25 residues) but are different in their acetylation state. H4K12Ac_H4K16 peptide has only acetylated K12 lysine and H4K12_H4K16Ac has only K16 acetylated. So both peptides have to bind to bromodomain alone due to presence of at least one epigenetic mark. This "control" weak binding is clearly seen without the reprogramming compound ('-' column) on the gel. On the other hand, both peptides have two unmodified lysines which are not recognized by bromodomain but should contribute to the binding upon reprogramming with the variator. Indeed, addition of NSC689846 dramatically increases intensively of bands on the gel ('+' column). Numerical densitometry shows approximately 10 folds stronger band in the case of H4K12Ac_H4K16 peptide binding and twice as stronger band in the case of H4K12_H4K16Ac binding. Besides, the similar modulating effect on binding of two different peptides is considered as an independent replication of the reprogramming experiment. The pulldowns thus ultimately indicates that the bromodomain variator enforces binding of bromodomain to its unnatural substrate presumably by changing physicochemical microenvironment of the binding site by chelating of lysine sidechain cation in hydrophobic environment of bromodomain. NSC689846 illustrates the changes via significantly increased binding to both partially acetylated peptides that harbor two cations in sequence of the peptide fragment. The pulldown results provide strong evidence in favor of the hypothesis of a triple complex formation in close proximity from BPTF binding site as the mechanism of variators acting.

*Discussion.*

Here we provide the second proof-of-principle example that we can synthetically reprogram the affinity of proteins to unnatural substrates. In particular, the artificial functional linkage between BPTF bromodomain and unacetylated states of H4K12, H4K16 and H4K20 were created. The reprogramming compound — variator — modify the normal activity of the protein and provide novel drug design opportunities similar to previously used PHD domain reprogramming for compensating for loss-of-functions MLL2 mutations. Of a special note is the fact of computational identification of reprogramming compounds in both reports. In either case the database was roughly pre-screened with rapid but inaccurate molecular docking approach and then promising compounds were intensively studied with visual inspection followed by simulations of molecular dynamics. Due to absence of adequate modeling interconnection between MD and docking approaches, the described protocols requires visual inspections of thousands of complexes and hundreds of trajectories which is overwhelming when it comes to scale-up. Authors hope that described success would stimulate wider involvement of the combined approach into drug design process which in turn results in new software applications for efficient combination of the separate methodologies in a sort of automated way.

General similarity of bromodomains from different proteins [30-32] can explain much lower (approximately 3 order of magnitude) effective dosage of bromodomain variators, comparing to previously reported PHD domain variators, due to reprogramming of several targets at the time. In the case of PHD domain our best hit showed only c.a. 180% of binding straightening in pulldowns meanwhile bromodomain variators have bigger reprogramming effect of 200-1000%. Though the latter difference should be considered with care as from avidity about 3 (there are three reprogramming motifs of approximately the same $K_d$ within H4 tail: H4K12, H4K16 and H4K20) only one — H4K16 — is known for its cooperative binding together with H3K4 methylation marks. Nevertheless, rather big differences in IC50 values from cell-based assays of PHD domain variators (low uM range) and bromodomain variators (low nM range) didn't allow us to reliably detect the cooperativity vs additivity between two reprogramming compounds against tumor cell lines. The potential cooperativity follows from BPTF co-localization at doubly-modified nucleosomes but only dominating additive effect is clearly seen in two-dimensional titrations (data not shown).

It is interesting that antitumor (and antiviral) activity of the identified scaffold was previously reported [33]. Although the study didn't claim explicit mechanism of action, the claimed structure-activity relationships (SARs) are consistent with our reprogramming hypothesis. The compounds does have wide and diverse net effect over the cell and virus infection essentially regardless to the type of viral infection that much better fits theory of reprogramming of core epigenetic reader than an assumption about specific inhibition of a certain protein or pathway. Besides the main SAR features from the report is importance of hydrophobic and bulky substitutions at NH2 terminus of the compound can be explained by locking role to the fragment that encloses chelating imine reprogramming motif in active site of bromodomain in our theoretical model of triple reprogrammed complex.

*Conclusions.*

The computational rational design of conceptually new class of drugs — variators — is replicated for the second domain, bromodomain, of BPTF protein. Variators, as opposed to widely developed inhibitors, do not inhibit existing biochemical processes but create new artificial pathways *in vivo* by instigating novel molecular interactions on reprogrammed proteins. With this approach we were able to provide the second example of modified protein selectivity by forcing BPTF to bind to its new unnatural substrate – unmodified histone H4 peptide tails.

*Acknowledgements.*

Authors thank Dr Patel of Memorial Sloan-Kettering Cancer Center, New York, and Dr. C. David Allis of the Laboratory of Chromatin Biology, Rockefeller University for the full-length human BPTF construct.

Authors are grateful to the Developmental Therapeutics Program, National Cancer Institute, USA, for providing the chemical compounds in this project.

*Figures and Legends*

**Fig 1. Reprogramming effect of bromodomain variator NSC689846.**

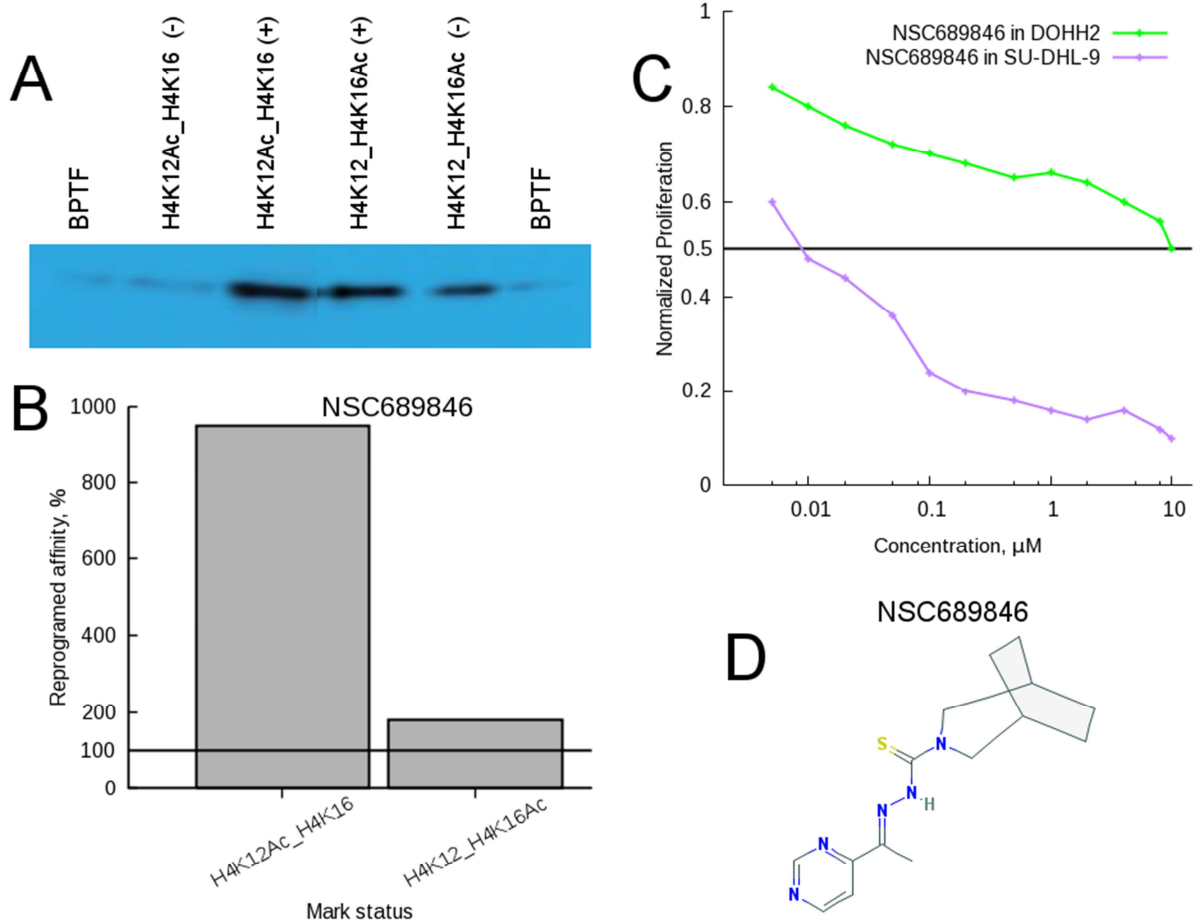

**A.** Pulldowns of NSC689846 with histone H4 N-terminal fragment residues 1-25. H4K12Ac_H4K16 is omodified by acetylation of K12 residue and H4K12_H4K16Ac is modified by acetylation of K16 residue, all other residues are unmodified in both cases. Plus (+) and minus (—) means presence and absence of the variators correspondingly. **B.** Relative response in % of densitometry units upon compound addition. **C.** Variator effect in cellular *in vitro* tests. **D.** Chemical structure of the variators NSC689846.